\documentclass[conference]{IEEEtran}

\ifCLASSINFOpdf
\usepackage{ifpdf}

\usepackage{cite}

\ifpdf
  \usepackage[pdftex]{graphicx}
  \DeclareGraphicsExtensions{.pdf}
\else
  \usepackage[dvips]{graphicx}
  \graphicspath{{../Figures/EPS/}}
  \DeclareGraphicsExtensions{.eps}
\fi

\usepackage[cmex10]{amsmath}

\usepackage{algorithm}
\usepackage{algorithmic}
\usepackage{array}

\usepackage{subfigure}
\usepackage{fixltx2e}
\usepackage{MnSymbol}
\usepackage{wasysym}
\usepackage{multirow}
\usepackage{url}

\newtheorem{defn} {Definition}

\newtheorem{prop}{Proposition}
\newcommand{\wusat}{%
\mathrel{\ooalign{$\Box$\kern-6.5pt$\filleddiamond$}}}

\begin{document}
%
\title{Two-Bit Bit Flipping Decoding of LDPC Codes}

\author{\IEEEauthorblockN{Dung Viet Nguyen, Bane Vasi$\acute{\mathrm{c}}$ and Michael W. Marcellin}
\IEEEauthorblockA{Department of Electrical and Computer Engineering\\
University of Arizona\\
Tucson, Arizona 85721\\
Email: \{nguyendv, vasic, marcellin\}@ece.arizona.edu}
}

\maketitle
\begin{abstract}
In this paper, we propose a new class of bit flipping algorithms for low-density parity-check (LDPC) codes over the binary symmetric channel (BSC). Compared to the regular (parallel or serial) bit flipping algorithms, the proposed algorithms employ one additional bit at a variable node to represent its ``strength.'' The introduction of this additional bit increases the guaranteed error correction capability by a factor of at least 2. An additional bit can also be employed at a check node to capture information which is beneficial to decoding. A framework for failure analysis of the proposed algorithms is described. These algorithms outperform the Gallager A/B algorithm and the min-sum algorithm at much lower complexity. Concatenation of two-bit bit flipping algorithms show a potential to approach the performance of belief propagation (BP) decoding in the error floor region, also at lower complexity. 
\end{abstract}

\section{Introduction}\label{sect_introduction}
High speed communication systems such as flash memory, optical communication and free space optics require extremely fast and low complexity error correcting schemes. Among existing decoding algorithms for LDPC codes \cite{ldpcBook_gallager} on the BSC, the bit flipping (serial or parallel) algorithms are least complex yet possess desirable error correcting abilities. First described by Gallager \cite{ldpcBook_gallager}, the parallel bit flipping algorithm was shown by Zyablov and Pinsker \cite{zyablov} to be capable of asymptotically correcting a linear number of errors (in the code length) for almost all codes in the regular ensemble with left-degree $\gamma\geq 5$. Later, Sipser and Spielman \cite{sipser} used expander graph arguments to show that this algorithm and the serial bit flipping algorithm can correct a linear number of errors if the underlying Tanner graph is a good expander. Note that their arguments also apply for regular codes with left-degree $\gamma\geq 5$. It was then recently shown by Burshtein \cite{burshtein} that regular codes with left-degree $\gamma=4$ are also capable of correcting a linear number of errors under the parallel bit flipping algorithm.

Despite being theoretically valuable, the above-mentioned capability to correct a linear number of errors is not practically attractive. This is mainly because the fraction of correctable errors is extremely small and hence the code length must be large. Besides, the above-mentioned results do not apply for column-weight-three codes, which allow very low decoding complexity. Also, compared to hard decoding message passing algorithms such as the Gallager A/B algorithm, the error performance of the bit flipping algorithms on finite length codes is usually inferior. This drawback is especially visible for column-weight-three codes for which the guaranteed error correction capability is upper-bounded by $\lceil g/4 \rceil-1$ (to be discussed later), where $g$ is the girth of a code. The fact that a code with $g=6$ or $g=8$ can not correct certain error patterns of weight two indeed makes the algorithm impractical regardless of its low complexity.

In recent years, numerous bit-flipping-oriented decoding algorithms have been proposed (see \cite{luckyCite} for a list of references). However, almost all of these algorithms require some soft information from a channel with capacity larger than that of the BSC. A few exceptions include the probabilistic bit flipping algorithm (PBFA) proposed by Miladinovic and Fossorier \cite{miladinovic}. In that algorithm, whenever the number of unsatisfied check nodes suggests that a variable (bit) node should be flipped, it is flipped with some probability $p<1$ rather than being flipped automatically. This random nature of the algorithm slows down the decoding, which was demonstrated to be helpful in practical codes whose Tanner graphs contain cycles. The idea of slowing down the decoding can also be found in a bit flipping algorithm proposed by Chan and Kschischang \cite{chan}. This algorithm, which is used on the additive white Gaussian noise channel (AWGNC), requires a certain number of decoding iterations between two possible flips of a variable node.

In this paper, we propose a new class of bit flipping algorithms for LDPC codes on the BSC. These algorithms are designed in the same spirit as the class of finite alphabet iterative message passing algorithms \cite{shiva}. In the proposed algorithms, an additional bit is introduced to represent the strength of a variable node. Given a combination of satisfied and unsatisfied check nodes, the algorithm may reduce the strength of a variable node before flipping it. An additional bit can also be introduced at a check node to indicate its reliability. The novelty of these algorithms is three-fold. First, similar to the above-mentioned PBFA, our class of algorithms also slows down the decoding. However they only do so when necessary and in a deterministic manner. Second, their deterministic nature and simplicity allow simple and thorough analysis. All subgraphs up to a certain size on which an algorithm fails to converge can be found by a recursive algorithm. Consequently, the guaranteed error correction capability of a code with such algorithms can be derived. Third, the failure analysis of an algorithm gives rise to better algorithms. More importantly, it leads to decoders which use a concatenation of two-bit bit flipping algorithms. These decoders show excellent trade offs between complexity and performance.

The rest of the paper is organized as follows. Section \ref{sect_pre} provides preliminaries. Section \ref{sect_tbd} motivates and describes the class of two-bit bit flipping algorithms. Section \ref{sect_analysis} gives a framework to analyze these algorithms. Finally, numerical results are presented in Section \ref{sect_numerical} along with discussion.

\section{Preliminaries}\label{sect_pre}
%
Let $\mathcal{C}$ denote an ($n,k$) LDPC code over the binary field GF(2). $\mathcal{C}$ is defined by the null space of $H$, an $m\times n$ \textit{parity check matrix}. $H$ is the bi-adjacency matrix of $G$, a Tanner graph representation of $\mathcal{C}$. $G$ is a bipartite graph with two sets of nodes: $n$ variable nodes and $m$ check nodes. In a $\gamma$-left-regular code, all variable nodes have degree $\gamma$. Each check node imposes a constraint on the neighboring variable nodes. A check node is said to be satisfied by a setting of variable nodes if the modulo-two sum of its neighbors is zero, otherwise it is unsatisfied. A vector ${\bf v} = \{v_1,v_2,\ldots,v_n\}$ is a codeword if and only if all check nodes are satisfied. The length of the shortest cycle in the Tanner graph $G$ is called the girth $g$ of $G$.

In this paper, we consider 3-left-regular LDPC codes with girth $g=8$, although the class of two-bit bit flipping algorithms can be generalized to decode any LDPC code. We assume transmission over the BSC. A variable node is said to be corrupt if it is different from its original sent value, otherwise it is correct. Throughout the paper, we also assume without loss of generality that the all-zero codeword is transmitted. Let ${\bf y} = \{y_1,y_2,\ldots,y_n\}$ denote the input to an iterative decoder. With the all-zero codeword assumption, the support of $\bf y$, denoted as $\mathrm{supp}({\bf y})$ is simply the set of variable nodes initially corrupt. In our case, a variable node is corrupt if it is 1 and is correct if it is 0. 

A simple hard decision decoding algorithm for LDPC codes on the BSC, known as the parallel bit flipping algorithm \cite{zyablov, sipser} is defined as follows. For any variable node $v$ in a Tanner graph $G$, let $n_c^{(s)}(v)$ and $n_c^{(u)}(v)$ denote the number of satisfied check nodes and unsatisfied check nodes that are connected to $v$, respectively. 
\begin{algorithm}
\caption{Parallel Bit Flipping Algorithm}
\begin{itemize}
\item In parallel, flip each variable node $v$ if $n_c^{(u)}(v)>n_c^{(s)}(v)$.
\item Repeat until all check nodes are satisfied.
\end{itemize}
\end{algorithm}
\section{The class of two-bit bit flipping algorithms} \label{sect_tbd}
The class of two-bit bit flipping algorithms is described in this section. We start with two motivating examples. The first one illustrates the advantage of an additional bit at a variable node while the second illustrates the advantage at a check node.
\subsection{First Motivating Example: Two-bit Variable Nodes}
\begin{figure}
\centering
\subfigure[] 
{
    \label{fp13}
\includegraphics[]{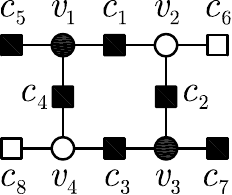}
}
\hspace{0.05in}
\subfigure[] 
{
    \label{fp24}
\includegraphics[]{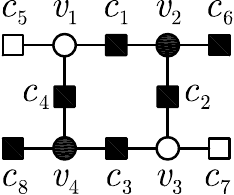}
}
\hspace{0.05in}
\subfigure[] 
{
    \label{fp12}

\includegraphics[]{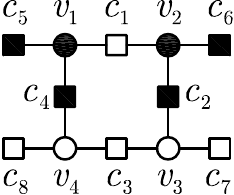}
}
\caption{Weight-two error configurations uncorrectable by the parallel bit flipping algorithm.}
\label{fig_fp}
\end{figure}

In this subsection, symbols $\fullmoon$ and $\newmoon$ denote a correct and a corrupt variable node while $\square$ and $\blacksquare$ denote a satisfied and an unsatisfied check node. 
Let $\mathcal{C}$ be a 3-left-regular LDPC code with girth $g=8$ and assume that the variable nodes $v_1, v_2, v_3$ and $v_4$ form an eight cycle as shown in  Fig.~\ref{fig_fp}. Also assume that only $v_1$ and $v_3$ are initially in error and that the parallel bit flipping algorithm is employed. In the first iteration illustrated in  Fig.~\ref{fig_fp}\subref{fp13}, $c_1,c_2,c_3,c_4,c_5$ and $c_7$ are unsatisfied while $c_6$ and $c_8$ are satisfied. Since $n_c^{(u)}(v_1) = n_c^{(u)}(v_3) = 3$ and $n_c^{(s)}(v_1) = n_c^{(s)}(v_3) = 0$, $v_1$ and $v_3$ are flipped and become correct. However, $v_2$ and $v_4$ are also flipped and become incorrect since $n_c^{(u)}(v_2) = n_c^{(u)}(v_4)= 2$ and $n_c^{(s)}(v_2) = n_c^{(s)}(v_4) = 1$. In the second iteration (Fig.~\ref{fig_fp}\subref{fp24}), the algorithm again flips $v_1, v_2, v_3$ and $v_4$. It can be seen that the set of corrupt variable nodes alternates between $\{v_1,v_3\}$ and $\{v_2,v_4\}$, and thus the algorithm does not converge. 

The parallel bit flipping algorithm fails in the above situation because it uses the same treatment for variable nodes with $n_c^{(u)}(v)= 3$ and $n_c^{(u)}(v)= 2$. The algorithm is too ``aggressive'' when flipping a variable node $v$ with $n_c^{(u)}(v)= 2$. Let us consider a modified algorithm which only flips a variable node $v$ with $n_c^{(u)}(v)= 3$. This modified algorithm will converge in the above situation. However, if only $v_1$ and $v_2$ are initially in error (Fig.~\ref{fig_fp}\subref{fp12}) then the modified algorithm does not converge because it does not flip any variable node. The modified algorithm is now too ``cautious'' to flip a variable node $v$ with $n_c^{(u)}(v)= 2$.

Both decisions (to flip and not to flip) a variable node $v$ with $n_c^{(u)}(v)= 2$ can lead to decoding failure. However, we must pick one or the other due the assumption that a variable node takes its value from the set $\{0,1\}$. Relaxing this assumption is therefore required for a better bit flipping algorithm.

Let us now assume that a variable node can take four values instead of two. Specifically, a variable node takes its value from the set $\mathcal{A}_v = \{0_s,0_w,1_w,1_s\}$, where $0_s$ ($1_s$) stands for ``strong zero'' (``strong one'') and $0_w$ ($1_w$) stands for ``weak zero'' (``weak one''). Assume for now that a check node only sees a variable node either as 0 if the variable node is $0_s$ or $0_w$, or as 1 if the variable node is $1_s$ or $1_w$. Recall that $n_c^{(u)}(v)\in\{0,1,2,3\}$ is the number of unsatisfied check nodes that are connected to the variable node $v$. Let $f_1:\mathcal{A}_v\times \{0,1,2,3\}\rightarrow \mathcal{A}_v$ be the function defined in Table \ref{tb_f1}.

\begin{table}[htb]
\caption{$f_1:\mathcal{A}_v\times \{0,1,2,3\}\rightarrow \mathcal{A}_v$}
\begin{center}
\begin{tabular}{||c|c|c|c|c||c|c|c|c|c||}\hline
\multirow{2}{*}{$v$}  &\multicolumn{4}{c||}{$n_c^{(u)}(v)$}&\multirow{2}{*}{$v$}  &\multicolumn{4}{c||}{$n_c^{(u)}(v)$}\\\cline{2-5}\cline{7-10}
 						 &0     &1     &2     &3  & &0     &1     &2     &3   \\\hline\hline
$0_s$				 &$0_s$ &$0_s$ &$0_w$	&$1_s$ & $1_w$				 &$1_s$ &$0_w$ &$0_s$ &$0_s$ \\\hline
$0_w$				 &$0_s$	&$1_w$ &$1_s$	&$1_s$ & $1_s$				 &$1_s$	&$1_s$ &$1_w$	&$0_s$\\\hline
\end{tabular}
\end{center}
\label{tb_f1}
\end{table}
\pagebreak

Consider the following bit flipping algorithm.
\begin{algorithm}[h]
\caption{Two-bit Bit Flipping Algorithm 1 (TBFA1)}
Initialization: Each variable node $v$ is initialized to $0_s$ if $y_v=0$ and is initialized to $1_s$ if $y_v=1$.
\begin{itemize}
\item In parallel, flip each variable node $v$ to $f_1(v,n_c^{(u)}(v))$.
\item Repeat until all check nodes are satisfied.
\end{itemize}
\end{algorithm}

Compared to the parallel bit flipping algorithm and its modified version discussed above, the TBFA1 possesses a gentler treatment for a variable node $v$ with $n_c^{(u)}(v)=2$. It tries to reduce the ``strength'' of $v$ before flipping it. One may realize at this point that it is rather imprecise to say that the TBFA1 flips a variable node $v$ from $0_s$ to $0_w$ or vice versa, since a check node still sees $v$ as 0. However, as the values of $v$ can be represented by two bits, i.e., $\mathcal{A}_v$ can be mapped onto the alphabet $\{01, 00, 10, 11\}$, the flipping of $v$ should be understood as either the flipping of one bit or the flipping of both bits.  

It is easy to verify that the TBFA1 is capable of correcting the error configurations shown in  Fig.~\ref{fig_fp}. Moreover, the guaranteed correction capability of this algorithm is given in the following proposition.
\begin{prop}\label{algo1cap}
The TBFA1 is capable of correcting any error pattern with up to $g/2-1$ errors in a left-regular column-weight-three code with Tanner graph $G$ which has girth $g\leq12$ and which does not contain any codeword of weight $w<g$.
\end{prop} 
\IEEEproof The proof is omitted due to page limits.
\endIEEEproof

\textit{Remarks:} 
\begin{itemize}
\item It can be shown that the guaranteed error correction capability of a 3-left-regular code with the parallel bit flipping algorithm is strictly less than $\lceil\frac{g}{4}\rceil$. Thus, the TBFA1 increases the guaranteed error correction capability by a factor of at least 2.
\item In \cite{col3Part2}, we have shown that the Gallager A/B algorithm is capable of correcting any error pattern with up to $g/2-1$ errors in a 3-left-regular code with girth $g\geq 10$. For codes with girth $g=8$ and minimum distance $d_{min}>8$, the Gallager A/B algorithm can only correct up to two errors. This means that the guaranteed error correction capability of the TBFA1 is at least as good as that of the Gallager A/B algorithm (and better for codes with $g=8$). It is also not difficult to see that the complexity of the TBFA1 is much lower than that of the Gallager A/B algorithm.
\end{itemize}

Now that the advantage of having more than one bit to represent the values of a variable node is clear, let us explore the possibility of using more than one bit to represent the values of a check node in the next subsection.
\subsection{Second Motivating Example: Two-bit Check Nodes}
\begin{figure}
\centering
\subfigure[] 
{
    \label{fvp1}

\includegraphics[]{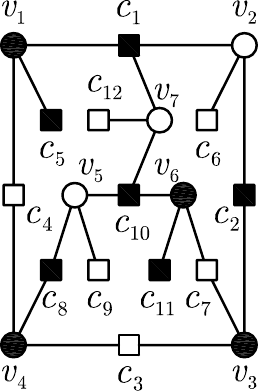}
}
\hspace{-0.065in}
\subfigure[] 
{
    \label{fvp2}
\includegraphics[]{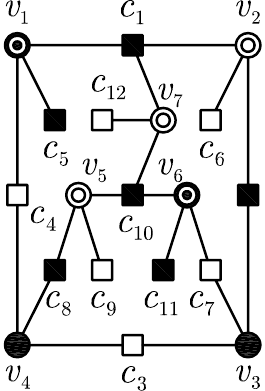}
}
\hspace{-0.065in}
\subfigure[] 
{
    \label{fvp3}
\includegraphics[]{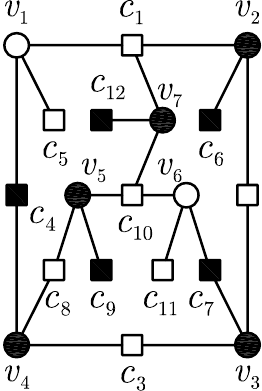}
}
\caption{The decoding of the two-bit parallel bit flipping algorithm 1 on a weight-four error configuration.}
\label{fig_fvp}
\end{figure}

In this subsection, we use the symbols $\fullmoon$ and $\newmoon$ to denote a $0_s$ variable node and a $1_s$ variable node, respectively. The symbols used to denote a $0_w$ variable node and a $1_w$ variable node are shown in Fig. \ref{fig_fvp}\subref{fvp2} where $v_2$ is a $0_w$ variable node and $v_1$ is a $1_w$ variable node . The symbols $\square$ and $\blacksquare$ still represent a satisfied and an unsatisfied check node.

Assume a decoder that uses the TBFA1 algorithm. Fig.~\ref{fig_fvp}\subref{fvp1}, \subref{fvp2} and \subref{fvp3} illustrates the first, second and third decoding iteration of the TBFA1 on an error configuration with four variable nodes $v_1,v_3,v_4$ and $v_6$ that are initially in error. We assume that all variable nodes which are not in this subgraph remain correct during decoding and will not be referred to. In the first iteration, variable nodes $v_1,v_2,v_5,v_6$ and $v_7$ are strong and connected to two unsatisfied check nodes. Consequently, the TBFA1 reduces their strength. Since variable nodes $v_3$ and $v_4$ are strong and only connected to one unsatisfied check node, their values are not changed. In the second iteration, all check nodes retain their values (satisfied or unsatisfied) from the first iteration. The TBFA1 hence flips $v_1$ and $v_6$ from $1_w$ to $0_s$ and flips $v_2,v_5$ and $v_7$ from $0_w$ to $1_s$. At the beginning of the third iteration, the value of any variable node is either $0_s$ or $1_s$. Every variable node is connected to two satisfied check nodes and one unsatisfied check node. Since no variable node can change its value, the algorithm fails to converge. 

The failure of the TBFA1 to correct this error configuration can be attributed to the fact that check node $c_3$ is connected to two initially erroneous variable nodes $v_3$ and $v_4$, consequently preventing them from changing their values. Let us slightly divert from our discussion and revisit the PBFA proposed by Miladinovic and Fossorier \cite{miladinovic}. The authors observed that variable node estimates corresponding to a number close to $\lceil \gamma/2\rceil$ unsatisfied check nodes are unreliable due to multiple errors, cycles in the code graph and equally likely a priori hard decisions. Based on this observation, the PBFA only flips a variable node with some probability $p<1$. In the above error configuration, a combination of two unsatisfied and one satisfied check nodes would be considered unreliable. Therefore, the PBFA would flip the corrupt variable nodes $v_1$ and $v_6$ as well as the correct variable node $v_2,v_5$ and $v_7$ with the same probability $p$. However, one can see that a combination of one unsatisfied and two satisfied check nodes would also be unreliable because such combination prevents the corrupt variable nodes $v_3$ and $v_4$ from being corrected. Unfortunately, the PBFA can not flip variable nodes with less than $\gamma/2$ unsatisfied check nodes since many other correct variable nodes in the Tanner graph would also be flipped. In other words, the PBFA can not evaluate the reliability of estimates corresponding to a number close to $\lfloor \gamma/2\rfloor$ unsatisfied check nodes. We demonstrate that such reliability can be evaluated with a new concept introduced below.

Revisit the decoding of the TBFA1 on the error configuration illustrated in Fig.~\ref{fig_fvp}. Notice that in the third iteration, except check node $c_3$, all check nodes that are unsatisfied in the second iteration become satisfied while all check nodes that are satisfied in the second iteration become unsatisfied. We will provide this information to the variable nodes. 

\begin{defn}\label{checkDef}
A satisfied (unsatisfied) check node is called \textit{previously satisfied} (\textit{previously unsatisfied}) if it was satisfied (unsatisfied) in the previous decoding iteration, otherwise it is called \textit{newly satisfied} (\textit{newly unsatisfied}). 
\end{defn}
\begin{figure}
\centering
\includegraphics[]{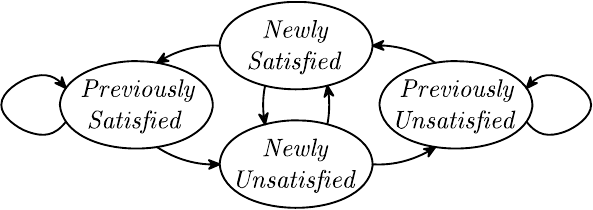}

\caption{Possible values and transition of a check node.}
\label{transCheck}
\end{figure}

The possible transitions of a check node are illustrated in Fig. \ref{transCheck}.
Let $n_c^{(s_p)}(v)$, $n_c^{(u_p)}(v)$, $n_c^{(s_n)}(v)$ and $n_c^{(u_n)}(v)$ be the number of previously satisfied check nodes, previously unsatisfied check nodes, newly satisfied check nodes and newly unsatisfied check nodes that are connected to a variable node $v$, respectively. Let $f_2:\mathcal{A}_v\times \{0,1,2,3\}^3\rightarrow \mathcal{A}_v$ be a function defined as follows:
\begin{eqnarray}
f_2(v,x,y,z) = f_1(v,x+y)\mathrm{~if~}(x,y,z)\notin \{(0,1,2),(0,1,1)\},\nonumber\\
f_2(v,0,1,2) = v,f_2(0_s,0,1,1) = f_2(0_w,0,1,1) = 0_w, \nonumber\\
f_2(1_s,0,1,1) = f_2(1_w,0,1,1) = 1_w.\nonumber
\end{eqnarray}

Consider the following bit flipping algorithm:
\begin{algorithm}
\caption{Two-bit Bit Flipping Algorithm 2 (TBFA2)}
Initialization: Each variable node $v$ is initialized to $0_s$ if $y_v=0$ and is initialized to $1_s$ if $y_v=1$. In the first iteration, check nodes are either previously satisfied or previously unsatisfied.
\begin{itemize}
\item In parallel, flip each variable node $v$ to \\$f_2(v,n_c^{(u_p)}(v),n_c^{(u_n)}(v),n_c^{(s_p)}(v))$.
\item Repeat until all check nodes are satisfied.
\end{itemize}
\end{algorithm}

The TBFA2 considers a combination of one newly unsatisfied, one newly satisfied and one previously satisfied check node to be less reliable than a combination of one previously unsatisfied and two previously satisfied check nodes. Therefore, it will reduce the strength of $v_3$ and $v_4$ at the end of the third iteration. Consequently, the error configuration shown in Fig. \ref{fig_fvp} can now be corrected after 9 iterations. Proposition \ref{algo1cap} also holds for the TBFA2. 

\textit{Remarks:} Let $\mathcal Q$ be the set of all functions from $\mathcal{A}_v\times \{0,1,2,3\}\rightarrow\mathcal{A}_v$. A natural question to ask is whether $f_1$ can be replaced with some $f'_1\in \mathcal{Q}$ such that the TBFA1 algorithm can correct the error configuration shown in Fig.~\ref{fig_fvp}. Brute force search reveals many of such functions. Unfortunately, none of those functions allow the algorithm to retain its guaranteed error correction capability stated in Proposition \ref{algo1cap}. 

We recap this section by giving the formal definition of the class of two-bit bit flipping algorithms.
\begin{defn}
For the class of two-bit bit flipping algorithms, a variable node $v$ takes its value from the set $\mathcal{A}_v = \{0_s,0_w,1_w,1_s\}$. A check node sees a $0_s$ and a $0_w$ variable node as 0 and sees a $1_s$ and a $1_w$ variable node as 1. According to Definition \ref{checkDef}, a check node can be previously satisfied, previously unsatisfied, newly satisfied or newly unsatisfied. An algorithm $\mathcal{F}$ is defined by a mapping $f:\mathcal{A}_v\times \{0,1,\ldots, \gamma\}^3\rightarrow \mathcal{A}_v$, where $\gamma$ is the column-weight of a code.
\end{defn}
%
%
%

Different algorithms in this class are specified by different functions $f$. In order to evaluate the performance of an algorithm, it is necessary to analyze its failures. To that task we shall now proceed.

\section{A framework for failure analysis}\label{sect_analysis}
In this section, we describe a framework for the analysis of two-bit bit flipping algorithms (the details will be provided in the journal version of this paper). Consider the decoding of a two-bit bit flipping algorithm $\mathcal{F}$ on a Tanner graph $G$. Assume a maximum number of $l$ iterations and assume that the channel makes $k$ errors. Let $I$ denote the subgraph induced by the $k$ variable nodes that are initially in error. Let $\mathcal{S}$ be the set of all Tanner graphs that contain $I$. Let $\mathcal{S}_e$ be the subset of $\mathcal{S}$ with the following property: if $S\in \mathcal{S}_e$ then there exists an induced subgraph $J$ of $S$ such that (i) $J$ is isomorphic to $I$ and (ii) the two-bit bit flipping algorithm $\mathcal{F}$ fails to decode on $S$ after $l$ iterations if the $k$ initially corrupt variable nodes are variable nodes in $J$. Let $\mathcal{S}_e^r$ be a subset of $\mathcal{S}_e$ such that any graph $S_1\in \mathcal{S}_e$ contains a graph $S_2\in \mathcal{S}_e^r$ and no graph in $\mathcal{S}_e^r$ contains another graph in $\mathcal{S}_e^r$. With the above formulation, we give the following proposition.
\begin{prop}\label{profF}
Algorithm $\mathcal{F}$ will converge on $G$ after $l$ decoding iterations if the induced subgraph $I$ is not contained in any induced subgraph $K$ of $G$ that is isomorphic to a graph in $\mathcal{S}_e^r$. 
\end{prop}
\IEEEproof If $\mathcal{F}$ fails to converge on $G$ after $l$ iterations then $G\in\mathcal{S}_e$, hence $I$ must be contained in an induced subgraph $K$ of $G$ that are isomorphic to a graph in $\mathcal{S}_e^r$.
\endIEEEproof
We remark that Proposition \ref{profF} only gives a sufficient condition. This is because $K$ might be contained in an induced subgraph of $G$ that is not isomorphic to any graph in $\mathcal{S}_e$. Nevertheless, $\mathcal{S}_e^r$ can still be used as a benchmark to evaluate the algorithm $\mathcal{F}$. A better algorithm should allow the above sufficient condition to be met with higher probability. For a more precise statement, we give the following.
\begin{prop}\label{profG}
The probability that a Tanner graph $I$ is contained in a Tanner graph $K_1$ with $k_1$ variable nodes is less than the probability that $I$ is contained in a Tanner graph $K_2$ with $k_2$ variable nodes if $k_1>k_2$
\end{prop}
\IEEEproof Let $K^{(s)}_2$ be a Tanner graph with $k_1$ variable nodes such that $K^{(s)}_2$ contains $K_2$. Since $K_1$ and $K^{(s)}_2$ both have $k_1$ variable nodes, the probability that $I$ is contained in $K_1$ equals the probability that $I$ is contained in $K^{(s)}_2$. On the other hand, since $K^{(s)}_2$ contains $K_2$, the probability that $I$ is contained in $K^{(s)}_2$ is less than the probability that $I$ is contained in $K_2$ by conditional probability.
\endIEEEproof
Proposition \ref{profG} suggests that a two-bit bit flipping algorithm should be chosen to maximize the size (in terms of number of variable nodes) of the smallest Tanner graph in $\mathcal{S}_e^r$. Given an algorithm $\mathcal{F}$, one can find all graphs in $\mathcal{S}_e^r$ up to a certain number of variable nodes by a recursive algorithm. Let $V_e^{(i)}$ denote the set of corrupt variable nodes at the beginning of the $i$-th iteration. The algorithm starts with the subgraph $I$, which is induced by the variable nodes in $V_e^{(1)}$. Let $\mathcal{N}(V_e^{(i)})$ be the set of check nodes that are connected to at least one variable node in $V_e^{(i)}$. In the first iteration, only the check nodes in $\mathcal{N}(V_e^{(1)})$ can be unsatisfied.  Therefore, if a correct variable node becomes corrupt at the end of the first iteration then it must connect to at least one check node in $\mathcal{N}(V_e^{(i)})$. In all possible ways, the algorithm then expands $I$ recursively by adjoining new variable nodes such that these variable nodes become corrupt at the end of the first iteration. The recursive introduction of new variable nodes halts if a graph in $\mathcal{S}_e^r$ is found. Let $\mathcal{T}_1$ be the set of graphs obtained by expanding $I$. Each graph in $\mathcal{T}_1\backslash\mathcal{S}_e^r$ is then again expanded by adjoining new variable nodes that become corrupt at the end of the second iteration. This process is repeated $l$ times where $l$ is the maximum number of iterations. 


\section{Numerical results and Discussion}\label{sect_numerical}
We demonstrate the performance of two-bit bit flipping algorithms on a regular column-weight-three quasi-cyclic LDPC code of length $n=768$. The code has rate $R=0.75$ and minimum distance $d_\mathrm{min}=12$. Two different decoders are considered. The first decoder, denoted as BFD1, employs a single two-bit bit flipping algorithm. The BFD1 may perform iterative decoding for a maximum number of 30 iterations. The second decoder, denoted as BFD2, is a concatenation of 55 algorithms, namely $\mathcal{F}_1, \mathcal{F}_2,\ldots\mathcal{F}_{55}$. Associated with algorithm $\mathcal{F}_i$ is a maximum number of iterations $l_i$. The BFD2 operates by performing decoding using algorithm $\mathcal{F}_i$ on an input vector $\bf y$ for $i=1,2,\ldots,55$ or until a codeword is found. The maximum possible number of decoding iterations performed by the BFD2 is $\sum_{i=1}^{55}{l_i} = 1950$. Details on the algorithms $\mathcal{F}_1, \mathcal{F}_2,\ldots\mathcal{F}_{55}$ as well as the parity check matrix of the quasi-cyclic LDPC code can be found in \cite{website}.

\begin{figure}
\centering
\includegraphics[width = 3.1in]{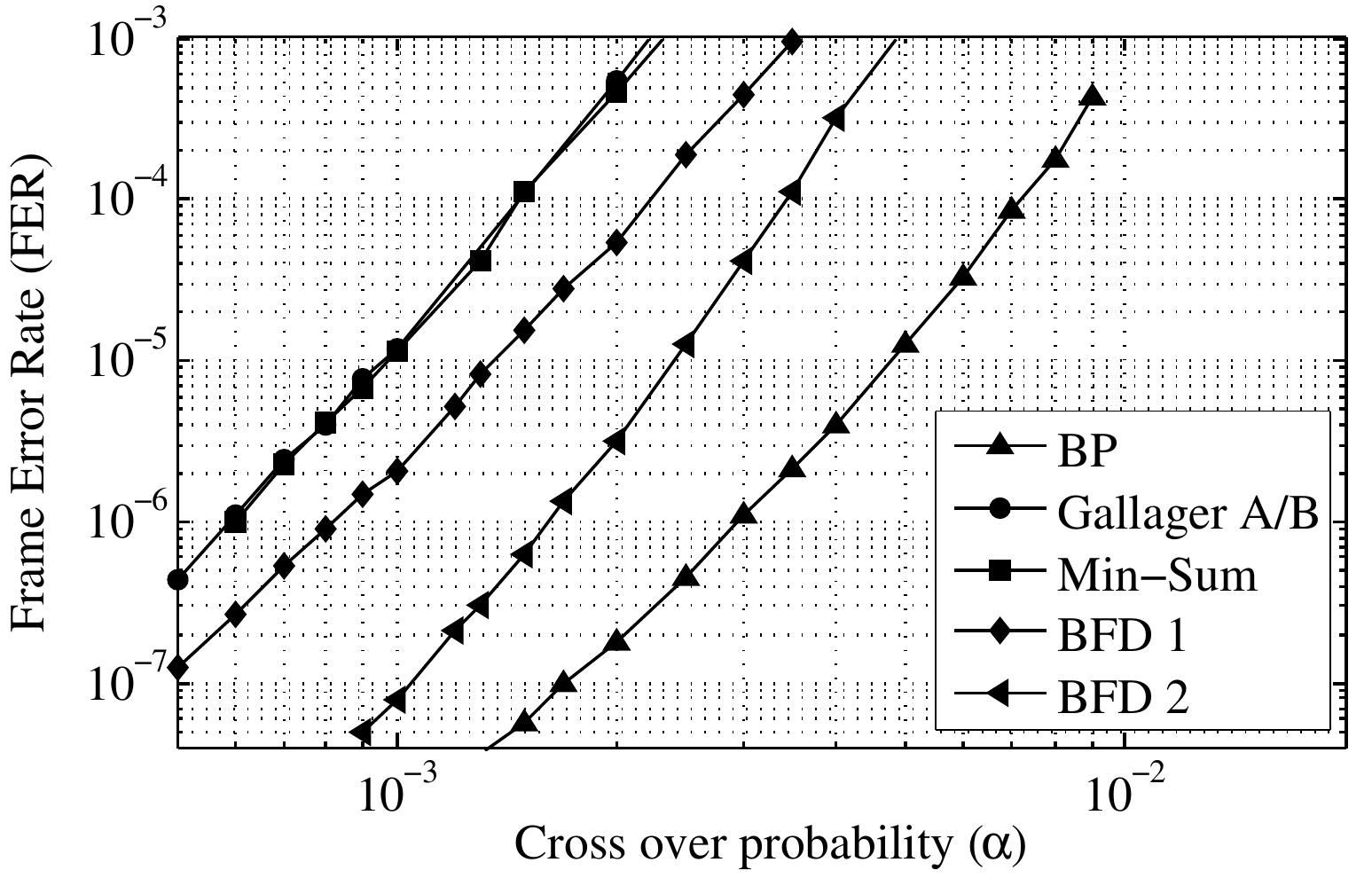}
\caption{Frame error rate performance of the BFD1 and BFD2.}
\label{fig_fer}
\end{figure}

\begin{figure}
\centering
\includegraphics[width = 3.1in]{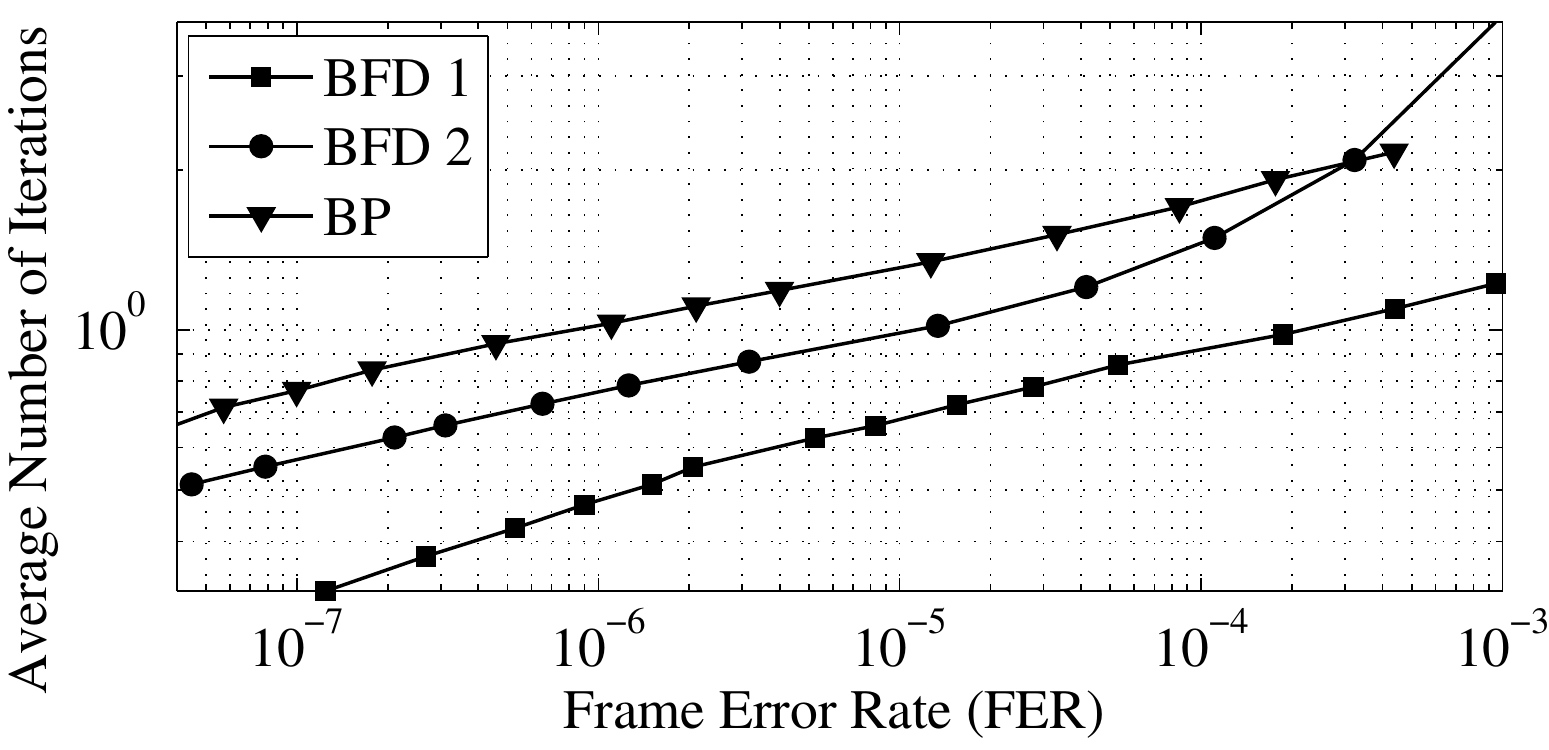}
\caption{Average number of decoding iterations per output word.}
\label{fig_iter}
\end{figure}

Simulations for frame error rate (FER) are shown in Fig. \ref{fig_fer}. Both decoders outperform decoders which use the Gallager A/B algorithm or the min-sum algorithm. In particular, the FER performance of the BFD2 is significantly better. More importantly, the slope of the FER curve of the BFD2 is larger than that of the BP decoder. This shows the potential of two-bit bit flipping decoders with comparable or even better error floor performance than that of the BP decoder. It is also important to remark that although the BFD2 uses 55 different decoding algorithms, at cross over probability $\alpha<0.0025$, more than 99.99\% of codewords are decoded by the first algorithm. Consequently, the average number of iterations per output word of the BFD2 is not much higher than that of the BFD1, as illustrated in Fig. \ref{fig_iter}. This means that similar to the BFD1, the BFD2 has an extremely high speed.



\section*{Acknowledgment}
This work was funded by NSF under the grants CCF-0963726 and CCF-0830245.


\begin{thebibliography}{10}
\providecommand{\url}[1]{#1}
\csname url@samestyle\endcsname
\providecommand{\newblock}{\relax}
\providecommand{\bibinfo}[2]{#2}
\providecommand{\BIBentrySTDinterwordspacing}{\spaceskip=0pt\relax}
\providecommand{\BIBentryALTinterwordstretchfactor}{4}
\providecommand{\BIBentryALTinterwordspacing}{\spaceskip=\fontdimen2\font plus
\BIBentryALTinterwordstretchfactor\fontdimen3\font minus
  \fontdimen4\font\relax}
\providecommand{\BIBforeignlanguage}[2]{{%
\expandafter\ifx\csname l@#1\endcsname\relax
\typeout{** WARNING: IEEEtran.bst: No hyphenation pattern has been}%
\typeout{** loaded for the language `#1'. Using the pattern for}%
\typeout{** the default language instead.}%
\else
\language=\csname l@#1\endcsname
\fi
#2}}
\providecommand{\BIBdecl}{\relax}
\BIBdecl

\bibitem{ldpcBook_gallager}
R.~G. Gallager, \emph{Low Density Parity Check Codes}.\hskip 1em plus 0.5em
  minus 0.4em\relax Cambridge, MA: M.I.T. Press, 1963.

\bibitem{zyablov}
V.~Zyablov and M.~Pinsker, ``Estimation of the error-correction complexity for
  {G}allager low-density codes,'' \emph{Probl. Inf. Transm.}, vol.~11, no.~6,
  pp. 18--26, 1976.

\bibitem{sipser}
M.~Sipser and D.~Spielman, ``Expander codes,'' \emph{IEEE Trans. Inf. Theory},
  vol.~42, no.~6, pp. 1710--1722, Nov. 1996.

\bibitem{burshtein}
D.~Burshtein, ``On the error correction of regular {LDPC} codes using the
  flipping algorithm,'' \emph{IEEE Trans. Inf. Theory}, vol.~54, no.~2, pp.
  517--530, Feb. 2008.

\bibitem{luckyCite}
X.~Wu, C.~Ling, M.~Jiang, E.~Xu, C.~Zhao, and X.~You, ``New insights into
  weighted bit-flipping decoding,'' \emph{IEEE Trans. Commun.}, vol.~57, no.~8,
  pp. 2177--2180, Aug. 2009.

\bibitem{miladinovic}
N.~Miladinovic and M.~Fossorier, ``Improved bit-flipping decoding of
  low-density parity-check codes,'' \emph{IEEE Trans. Inf. Theory}, vol.~51,
  no.~4, pp. 1594--1606, Apr. 2005.

\bibitem{chan}
A.~Chan and F.~Kschischang, ``A simple taboo-based soft-decision decoding
  algorithm for expander codes,'' \emph{IEEE Commun. Letters}, vol.~2, no.~7,
  pp. 183--185, Jul. 1998.

\bibitem{shiva}
S.~Planjery, D.~Declercq, S.~Chilappagari, and B.~Vasic~and, ``Multilevel
  decoders surpassing belief propagation on the binary symmetric channel,'' in
  \emph{IEEE Int. Symp. Inf. Theory}, Jun. 2010, pp. 769--773.

\bibitem{col3Part2}
S.~K. Chilappagari, D.~V. Nguyen, B.~V. Vasic, and M.~W. Marcellin, ``Error
  correction capability of column-weight-three {LDPC} codes under the
  {G}allager {A} algorithm - {P}art {II},'' \emph{IEEE Trans. Inf. Theory},
  vol.~56, no.~6, pp. 2626--2639, Jun. 2010.

\bibitem{website}
\BIBentryALTinterwordspacing
``Error floors of {LDPC} codes - {M}ulti-bit bit flipping algorithm.''
  [Online]. Available:
  \url{http://www2.engr.arizona.edu/~vasiclab/Projects/CodingTheory/ErrorFloor%
Home.html}
\BIBentrySTDinterwordspacing

\end{thebibliography}
\end{document}